\documentclass[twocolumn,prl,superscriptaddress,showpacs]{revtex4}

\usepackage{amsmath,amssymb,graphicx,textcomp}

\usepackage{avant}
\usepackage[T1]{fontenc}

\usepackage{amsmath,amssymb,graphicx}
\usepackage{color,textcomp}

\usepackage[normalem]{ulem}

\newcommand\delete[1]{}

\newcommand\diff{\mathrm{d}}

\usepackage{color}

\begin{document}

\title{Microscale fluid flow induced by thermoviscous expansion along a traveling wave}
\author{Franz M. Weinert}
\affiliation{Applied Physics and Center for NanoScience (CeNS),
Department of Physics, Ludwig-Maximilians-Universit{\"a}t
M{\"u}nchen, Amalienstrasse 54, 80799 M{\"u}nchen, Germany}
\author{Jonas A. Kraus}
\affiliation{Arnold Sommerfeld Center for Theoretical Physics
(ASC)  and Center for NanoScience (CeNS), Department of Physics,
Ludwig-Maximilians-Universit{\"a}t M{\"u}nchen,
Theresienstra{\ss}e 37, 80333 M{\"u}nchen, Germany}
\author{Thomas Franosch}
\affiliation{Arnold
Sommerfeld Center for Theoretical Physics (ASC)  and Center for
NanoScience (CeNS), Department of Physics,
Ludwig-Maximilians-Universit{\"a}t M{\"u}nchen, Theresienstra{\ss}e 37,
80333 M{\"u}nchen, Germany}
\author{Dieter  Braun}
\affiliation{Applied Physics and Center for NanoScience (CeNS),
Department of Physics, Ludwig-Maximilians-Universit{\"a}t
M{\"u}nchen, Amalienstrasse 54, 80799 M{\"u}nchen, Germany}

\begin{abstract}
The thermal expansion of a fluid combined with a
temperature-dependent viscosity introduces nonlinearities in the
Navier-Stokes equations unrelated to the convective momentum
current. The couplings generate the possibility for net fluid flow
at the microscale controlled by external heating. This novel
thermo-mechanical effect is investigated  for a thin fluid chamber
by a numerical solution of the Navier-Stokes equations and
analytically by a perturbation expansion. A demonstration
experiment confirms  the  basic mechanism  and quantitatively
validates our theoretical analysis.

\end{abstract}


\maketitle

Spatial confinement of a liquid changes its flow behavior
markedly, since the importance of surface forces relative to the
volume forces increases as the confinement becomes smaller.
Recently, flow at the scale of millimeters and below has attracted
significant attention, stimulated by the rapid advances to
manipulate and to control small-scale
devices~\cite{Stone:2004,Whitesides:2006,Squires:2005,Thorsen:2002}. Since microfluidic flow often
is essentially at zero Reynolds number, viscous drag overwhelms
the inertial effects of the fluid giving rise to peculiar flow
behavior~\cite{Happel:1983,Batchelor:1977}. In particular, heat
conduction becomes quite efficient, since the thermal diffusivity $\kappa = k/\rho c_p$, with $k$ the thermal conductivity and
$\rho c_p$ the volumetric heat capacity, implies
thermal relaxation times
$D^2/ \pi^2 \kappa$ of the order of $0.1 \mbox{ms}$ for water
confined by walls separated by $D = 10$\,\textmu{}m.  Usually, this strong
thermal coupling gives rise to a uniform temperature and all
physical processes are isothermal. Since pressure effects upon
density are negligible if the velocities are small with respect to
the speed of sound,  fluid flow is essentially "dynamically
incompressible" \cite{Batchelor:1967}, i.e. the velocity field is
solenoidal, $\mbox{div } \vec{v} =0$.

Recently, it has been noted by \textcite{Yariv:2004} that for the case
of unsteady heating this is no longer the case and one can in
principle generate a  non-solenoidal flow. A transient  fluid flow emerges
due to a thermal expansion of the fluid as a reaction to the
heating in combination with the thermal diffusion.

In this Letter we propose a novel mechanism to generate net  flow
in a thin fluid chamber, i.e. a viscous liquid confined between
two plates separated by a distance  of the order of a few
micrometers.  
The driving of the fluid flow is
provided  by imposing a traveling  temperature wave.  We  show analytically within a thin-film
approximation that such unsteady heating  leads to net fluid flow. Then we corroborate
our  analytic approach  by a finite-element calculation. Last we provide first experimental
evidence that there is indeed net flow and that the fundamental
dependences have been correctly identified.

The basic mechanism
may be summarized as follows: Due to  thermal expansion of the
liquid the motion of the heating results in a pressure modulation.
The pressure gradients induce a potential flow which, however,
does not lead to net fluid flow. Yet, the temperature dependence of the
shear viscosity gives rise to a small net mass transport typically
opposite to the motion of the heat source. The flow velocity is
then proportional to the thermal expansion coefficient of the
fluid, $\alpha = \rho_0^{-1} (\partial \rho/\partial T)$ as well
as to the thermal viscosity coefficient $\beta = - \eta_0^{-1}
(\partial \eta/\partial T)$.

The geometry under consideration consists of a thin chamber of
height $D$ and a much larger lateral extension. There is a natural
small parameter $\epsilon = D/L \ll 1$, where $L$ is a typical
length scale in the lateral direction.
 Since the film is thin one expects the Navier-Stokes equations to be dominated
only by a few terms. To identify the relevant contributions
dimensionless quantities are introduced as $ t = T t^*, v_i =
U v_i^*  ,  v_i = v v_\perp^*,x_i = L x_i^* ,
x_\perp = D x_\perp^* , p = P p^*$. Lateral directions are labeled
by a latin index $i\in \{ x, y \}$, whereas the vertical direction is indicated by
the subscript $\perp$. To keep all terms in the mass conservation
law the scales are chosen as $v = \epsilon U,  T = L/U$, and
consequently the fluid flow is essentially in-plane. To balance
the leading order term in the momentum conservation law the
pressure scale has to be chosen as $P = \eta_0 U/ (\epsilon^2
L)$. Momentum transport via convective processes may be ignored
from the very beginning since the Reynolds number
is small, $\mbox{\sl Re} = \rho_0 U L/\eta_0\ll 1$~\footnote{The  dimensionless number quantifying the relative importance of
convective to viscous terms for a thin film geometry is actually even
 smaller $\text{\sl Re}^* = \rho_0 U h^2/ L \eta_0$~\cite{Leal:1992}.}.
Then the mass conservation law and the momentum balance in
perpendicular and lateral direction read to leading order in a
\emph{thin-film approximation}, see e.g.
\cite{Obrien:2002,Oron:1997,Leal:1992} (restoring units)
\begin{subequations}
\begin{eqnarray}\label{mass_cons}
\partial_t \rho +  \nabla_\perp (\rho v_\perp) + \nabla_i ( \rho v_i) &= &  0 \, , \\
\label{pressure_perp} \nabla_\perp p &= &  0 \, , \\
\label{pressure_lateral} \nabla_\perp [\eta \nabla_\perp v_i]
 +\nabla_i p &= & 0 \, .
\end{eqnarray}
\end{subequations}
As a consequence of the thin film geometry,  (i) bulk viscous
processes do not contribute to the leading order equations, (ii)
the pressure is homogeneous in the perpendicular direction, (iii)
lateral pressure gradients drive the lateral fluid flow. Note that
the divergence of the velocity field is irrelevant for the
momentum balance but plays a crucial role in the mass conservation
law.

The coupling to the temperature enters the equations via the
expansion of the fluid as well as by the temperature dependence of
the shear viscosity.
 For the problem at hand it is appropriate to
neglect the mechanical compressibility of the fluid and to expand
the equation of state to first order in the temperature field,
$\rho = \rho_0 (1- \alpha \delta T)$, where $\delta T$ is the
local temperature change and $\alpha = -(\partial \ln
\rho/\partial T)_p$ the thermal expansion coefficient at the
reference state $(\rho_0,T_0)$. The density field then is
eliminated from the equations of motion in favor of the
temperature field $\delta T$. A second ingredient for the
theoretical model is to include the variation of the shear
viscosity with temperature. Introducing the thermal viscosity
coefficient $\beta = -(\partial \ln \eta/\partial T)_p$,  
the coupling reads $\eta = \eta_0 (1-\beta \delta T)$ to  first
order.

In the case of considerable thermal coupling to the walls the temperature
field approximately reflects the profile of the heat source and
then the temperature is uniform perpendicularly to the walls. The
in-plane variations of the temperature field occur on scales $\lambda, b
\gg D$, where $\lambda$ denotes the wavelength and $b$  the typical
lateral extension of the heating.
Consequently, the spatial dependence of the density and the shear
viscosity is only in-plane as it is inherited from the temperature
profile. With this additional assumption, the in-plane momentum
balance equation, Eq.~(\ref{pressure_lateral}),  is readily
integrated for
 no-slip boundary conditions at the walls 
\begin{eqnarray}
v_i = - \frac{1}{2\eta} z(D-z) \nabla_i p \, ,
\end{eqnarray}
i.e. the velocity profile corresponds to Poiseuille flow.

\begin{figure}
\includegraphics[width=\linewidth]{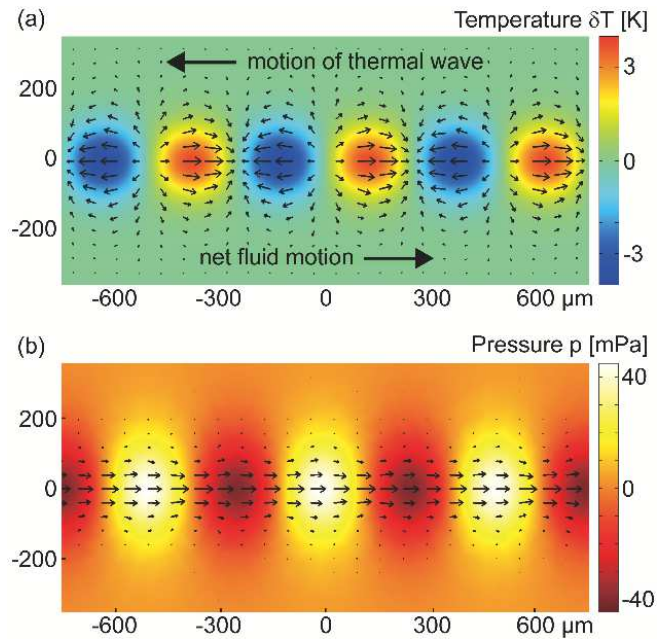}
\caption{(color online) Finite element solution for a  temperature wave in the comoving frame.
(a) The temperature profile is indicated by the color scale and the arrows refer to the induced velocity flow.
(b) The corresponding pressure modulation (color scale)   and the solenoidal component of the flow.} \label{fig:pressure}
\end{figure}

Since the spatial dependence of the velocity profiles
perpendicular to the wall is known, one may project the
three-dimensional problem to the two-dimensional plane by
averaging over the vertical $z$-direction, \begin{eqnarray}
\label{two_flow} \bar{v}_i = \frac{1}{D} \int_0^D v_i
\diff z = - \frac{D^2}{12 \eta}\nabla_i p\, . \end{eqnarray}
Because the density is assumed to vary only in the lateral direction, the  mass conservation law, Eq.~(\ref{mass_cons}), can be directly
averaged over the perpendicular direction, $\partial_t \rho +
\nabla_i (\rho \bar{v}_i) = 0$, without introducing new
terms. The pumping process becomes stationary in the frame of reference
comoving with the heat wave, i.e. substituting $\bar{v}_x \mapsto \bar{v}_x
-u$, where $u$ denotes the velocity of the heat wave. Then in the comoving frame the
averaged mass conservation law yields
\begin{eqnarray}\label{averaged_mass}
-u \nabla_x \rho +  \rho \nabla_i \bar{v}_i = -
\bar{v}_i \nabla_i \rho\, .
\end{eqnarray}
Here the velocities $\bar{v}_i$ are small quantities, typically
$\bar{v}_i \ll u$.
From the closed set of equations, Eqs.~(\ref{two_flow}) and (\ref{averaged_mass}),   
the fluid velocity can be calculated.  The temperature profile in
the comoving frame enters Eq.~(\ref{averaged_mass}) in the form of
density gradients, $\nabla_i \rho = - \alpha \rho_0
\nabla_i \delta T$ and Eq.~(\ref{two_flow}) via the
temperature dependent viscosity. The net flow emerges only
if the thermal viscosity coefficient~$\beta$ is non-vanishing. For
a temperature-independent viscosity  the pressure acts as a
velocity potential, Eq.~(\ref{two_flow}), and averaging along the
direction of the wave propagation yields a vanishing net flow.

We exemplify the
  flow profiles for  the case of a gaussian
temperature wave $\delta T(\vec{r}) = \delta T_0 \cos(k x)\,
\text{e}^{-y^2/2b^2}$,
obtained by a numerical solution of  Eqs.~(\ref{two_flow}) and (\ref{averaged_mass})  using the Femlab\textregistered 3.1 software
package by Comsol.
We have used a wavelength $\lambda=2\pi/k = 50$\,\textmu{}m, a
lateral
width $b=7$\,\textmu{}m of
the gaussian modulation, a temperature amplitude of $\delta T_0=3.8\,\text{K}$, and a thermal wave velocity $u=100$\,mm/s.
The parameters for
 $\alpha=3.0 \cdot
10^{-4}\,\text{K}^{-1}$, $\beta=0.022\,\text{K}^{-1}$ correspond to water at ambient  temperature.
The parameters have been chosen such that the temperature modulation appears
approximately as a train of circular heated areas mimicking 
the demonstration experiment below. The regions where the temperature gradient is maximal (minimal) act as sources (sinks) for the velocity field, giving rise to a locally dipolar pattern of the fluid flow,  see
Fig.~\ref{fig:pressure}(a). Careful inspection reveals
that the velocities in the heated regions are bigger than in the colder ones.
The pressure profile is shifted by a quarter of a wavelength reflecting the local expansion of the fluid, Fig.~\ref{fig:pressure}(b). The
solenoidal part of the velocity field has been obtained as the difference of two
Femlab calculations  for thermal viscosity coefficients $\beta \neq 0$ and $\beta = 0$.
Close to the center of the wave this contribution is  directed opposite to the thermal wave velocity
with
an average pump velocity $v_{\text{fl}}^{\text{FEM}}=2.88$\,\textmu{}m/s 
at the center of the thermal wave, whereas far away from the
heating  there is a characteristic  backflow. 
The pressures induced by the pumping are below $10^{-6}$
bar (Figure 1b). The pressure limitations of the glass chamber used are estimated
to 10~bar by a finite element calculation. However, the pressure induced by the pumping 
increases with the viscosity of the fluid
and eventually will stall the pump motion since thermal expansion
will deform only the chamber walls instead of triggering fluid motion. So we expect to equally
pump fluids with viscosities $10^7$ fold higher than water.


To gain further insight, we develop  a perturbative analytical solution  for the fluid flow
appropriate for small temperature changes. Formally, the expansion is performed in
the small parameters $\alpha$ and $\beta$, and we shall show that
net fluid flow  first occurs  at ${\cal O}(\alpha \cdot \beta)$. The
quiescent fluid $v_i \equiv 0, p = p_0$  solves the governing equations
to zeroth order in $\alpha, \beta$, i.e. if all couplings to the
temperature field are ignored. To leading order in $\alpha$, the
term on the r.h.s. of Eq.~(\ref{averaged_mass}) is of second order
and may be ignored. Similarly,  ~(\ref{two_flow}) is expanded
to first order in the temperature change
\begin{eqnarray}\label{beta_flow}
\bar{v}_i =  - \frac{D^2}{12 \eta_0 } (\nabla_i p) (
1+\beta \delta T) \, .
\end{eqnarray}
Already at this point one concludes that to order ${\cal
O}(\beta^0)$ the velocity corresponds to potential flow only,
hence no net pumping arises to that order. The leading order to
the pump velocity is then expected to be of order ${\cal O}(\alpha
\beta)$. To identify the net fluid motion it is favorable to
introduce the velocity potential, $\chi$, and the
stream function,$\psi$, via $\bar{v}_x = \nabla_x \chi + \nabla_y
\psi, \bar{v}_y =
 \nabla_y \chi - \nabla_x \psi$.
From Eq.~(\ref{beta_flow}) one infers that the leading term
corresponds to pure gradient flow, i.e. $\chi = -(D^2/12 \eta_0)
p$, where $\chi(\vec{r})$ denotes the  velocity potential and
$\vec{r} = (x,y)$. To the required order $\chi$ is determined by
combining Eqs. (\ref{averaged_mass})
\begin{eqnarray}\label{Poisson} \nabla^2 \chi(\vec{r}) = -\alpha u \nabla_x
\delta T(\vec{r}) +{\cal O}(\alpha^2) \, .
\end{eqnarray}
Once the velocity potential is determined, the 
stream function is calculated  by  extracting from $\beta
\delta T(\vec{r}) \nabla_i \chi(\vec{r}) $ its solenoidal component. By Helmholtz's decomposition
theorem, the stream function satisfies  a Poisson equation
\begin{eqnarray}\label{psi}
-\nabla^2 \psi = \nabla_x (\beta  \delta T  \nabla_y \chi) -
\nabla_y (\beta  \delta T  \nabla_x \chi)  \, .
\end{eqnarray}
The lowest order contribution to the stream function is thus
proportional to $\alpha \cdot \beta $, which also sets the overall
scale of the thermo-mechanical effect. Furthermore one infers from
 Eq.~(\ref{psi}), that a strictly one-dimensional temperature
modulation does not give rise to solenoidal flow. The remaining
task is to solve the set of Poisson equations, which can be easily
implemented by numerical methods. To determine the average flow
for a travelling periodic temperature wave train of wavelength
$\lambda$, it is actually sufficient to know the velocity
potential,
\begin{eqnarray}
v_\text{fl}(y) & \equiv &  \frac{1}{\lambda} \int_{0}^\lambda\!\!
\diff x \, \bar{v}_x(\vec{r})
  =  -\frac{1 }{\lambda} \int_{0}^\lambda\!\!\diff x\, \chi(\vec{r}) \beta \nabla_x \delta T(\vec{r}) \, ,\nonumber \\
\end{eqnarray}
where the last relation follows from the representation,
Eq.~(\ref{beta_flow}) and an integration by parts.

To illustrate the physics we consider a gaussian
temperature wave $\delta T(\vec{r}) = \delta T_0 \cos(k x)
\text{e}^{-y^2/2b^2}$, where  $b$ characterizes the
lateral width of the wave train. The Poisson equation, Eq.~(\ref{Poisson}), can be solved
 exactly in terms of error functions~\cite{Kraus:diploma}, however, it is instructive
to construct an approximate solution for wide waves, $k b \gg 1$.
Then the problem is effectively one-dimensional, and $\chi$
depends on the lateral coordinate $y$ only parametrically. One readily calculates  $\chi =- u \alpha
\delta T_0 \sin(k x) \text{e}^{-y^2/2b^2}/k$,
implying a net fluid motion
\begin{eqnarray}\label{vflow}
 v_{\text{fl}}(y) = - u\alpha \beta (\delta T_0)^2 \exp(-y^2/b^2)/2\, ,
\end{eqnarray}
typically opposite to the motion
of the traveling temperature wave. Since by assumption the changes
in the density and the viscosity are small, the net fluid motion is
much slower than the velocity of the temperature wave,
$v_\text{fl} \ll u$.

\begin{figure}
\includegraphics[width=1.0\linewidth]{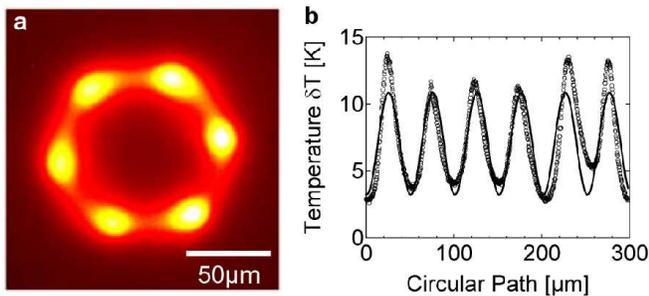}
\caption{(color online) Demonstration experiment. (a) A circular thermal wave is created in water by a moving
focus of an infrared laser. The temperature is inferred by stroboscopic imaging of a
temperature-sensitive fluorescent dye. (b)  Peak temperature along the
circumference. A sinusoidal fit
yields a temperature amplitude of $\delta T_0 =3.8\pm 0.5\,\text{K}$.}\label{fig:experiment}
\end{figure}

We have confirmed the predicted fluid movement in a demonstration
experiment. A circular variant of a thermal wave is imposed with
infrared light to a thin fluid film, see Fig.~\ref{fig:experiment}.
The fluid movement is recorded using micrometer-sized
fluorescent particles and the thermal wave is imaged
stroboscopically with temperature-sensitive fluorescence. Within
 experimental errors, the theory captures the thermally
triggered net flow.

The details of the experiment are as follows: A fiber laser at
1455 nm and a maximal power of 5 W (RLD-5-1455, IPGLaser) is
deflected by an acousto-optical deflector (Pegasus Optik,
AA.DTS.XY.100) and moderately focused from below (Thorlabs,
C240TM-C, $f= 8\,\text{mm}$, $\text{NA} = 0.5$) to a 10\,\textmu{}m
thin fluid film sandwiched between sapphire windows. The light is absorbed
by water with an attenuation length of 305\,\textmu{}m. The
chamber is imaged from above by a fluorescence microscope
(Zeiss, Axiotech Vario) and a CCD camera (SensiCam QE,
PCO). The illumination is provided by a green LED (LXHL-LX5C, Luxeon)
which for the temperature imaging was modulated with a bandwidth of
150\,kHz using a laser current source (LD-3565, ILX Lightwave).
The temperature field was imaged
using 50\,\textmu{}M of the fluorescent dye BCECF under stroboscopic illumination~\cite{Duhr:2004,Duhr:2006}
to determine the temperature amplitude of the thermal wave.

The thermal wave is generated by a scanning laser spot. Six individual
temperature peaks are created by a circular pattern of radius $R$=
50\,\textmu{}m with a base frequency of 30\,kHz. The frequency
is faster than the thermal relaxation time, 
which is dominated by the vertical heat currents and
determined to 0.21\,ms
using a finite element calculation. As result, a steady thermal
wave of six points is generated, measured in Fig.~\ref{fig:experiment}.
The points are rotated using a slower shift frequency $f_S= 0.2\ldots 2\,\text{kHz}$.
The result is a circular thermal wave with velocity $u= f_S \times 2 \pi R/6
= 10\ldots 100\,\text{mm/s}$.

\begin{figure}
\includegraphics[width=1.0\linewidth]{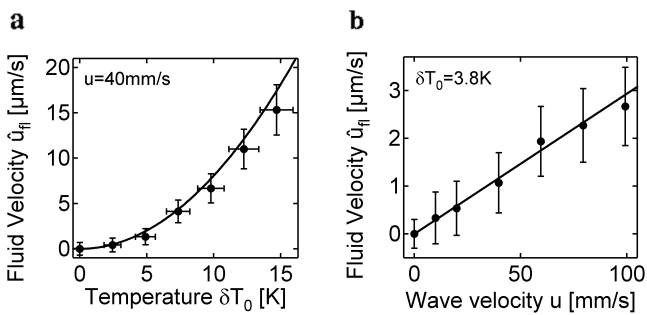}
\caption{(color online) Experiment versus theory. (a) The fluid velocity scales quadratically
with the amplitude of the thermal wave $\delta T_0$. (b) The speed increases linearly with
the velocity $u$ of the thermal wave. Measurements are given as dots, the theory as solid lines with the geometrical prefactor determined in
Ref.~\cite{Kraus:diploma}.
Error bars are standard errors.} \label{fig:measured}
\end{figure}

Fluid velocities are measured by single particle tracking of 80\,pM of 1\,\textmu{}m diameter silica beads (PSi-G1.0,
Kisker). Particle velocities could be well discriminated from Brownian diffusion by tracking 5
independent particles over 10\textmu{}m with a positional error of 1 \textmu{}m.
The peak velocity of the parabolic flow
profile was measured by selecting the beads in the chamber center.
The experimental values were multiplied by $2/3$ to compare
with the chamber-averaged theoretical values. Optical trapping or thermophoresis would move
the tracer particles with the thermal wave, opposite to the observed flow direction. Under
the experimental conditions, both effects would yield similar attractive forces. However, control experiments
in heavy water with its 100-fold decreased absorptive heating only revealed
moderate attraction into the illuminated ring, but no pumping movement along it (<0.13
\textmu{}m/s). This
indicates that the particles are ideal tracers of the fluid motion at the used focus size~\cite{Duhr:2006}.

With the above described temperature pattern and a thermal wave
velocity of $u= 100\,\text{mm/s}$ we measure a peak fluid flow of
$v_\text{fl}^{\text{exp}} = 2.7 \pm 0.5$\,\textmu{}m/s. We expect from
the wide wave solution, Eq.~\ref{vflow}, a fluid velocity of
$v_{\text{fl}}^{\text{theo}} = 4.77$\,\textmu{}m/s. Applying the complete analytical solution~\cite{Kraus:diploma} or the Femlab result yields $v_\text{fl}^{\text{FEM}} = 2.88$\,\textmu{}m/s which describes the
experimental result quantitatively.

Two parameter variations further confirm the theoretical model
(Fig.~\ref{fig:measured}). The dependence on the temperature
amplitude $\delta T_0$ exhibits a parabolic dependence
(Fig.~\ref{fig:measured}~a) as expected theoretically by Eq.~(\ref{vflow}). Furthermore, the fluid velocity scales linearly
with the velocity $u$ (Fig.~\ref{fig:measured}~b) of the thermal
wave as predicted (solid lines). We therefore find experimentally
that a circular thermal wave triggers a fluid flow according to
the theoretical description.

To conclude, we showed that a thermal wave can move a fluid by the nonlinear combination of the
temperature-dependent density and viscosity. While the obtainable fluid velocities are small
at present, future improvements could allow microfluidic applications.

\acknowledgements

We thank Joseph Egger for initial discussions. F.M.W. and D.B. were
supported by the Emmy Noether Scholarship of the DFG. Support by the
Nanosystems Initiative Munich (NIM) is gratefully acknowledged.


\end{document}